# Move As You Like: Image Animation in E-Commerce Scenario

Borun Xu[1], Biao Wang[2], Jiale Tao[1], Tiezheng Ge[2] ✉, Yuning Jiang[2], Wen Li[1], Lixin Duan[1]

[1]University of Electronic Science and Technology of China, [2]Alibaba Group

xbr_2017@std.uestc.edu.cn,{jialetao.std,liwenbnu}@gmail.com,{eric.wb,tiezheng.gtz,mengzhu.jyn,}@alibaba-inc.com,lxduan@uestc.edu.cn

## ABSTRACT

Creative image animations are attractive in e-commerce applications, where motion transfer is one of the import ways to generate animations from static images. However, existing methods rarely transfer motion to objects other than human body or human face, and even fewer apply motion transfer in practical scenarios. In this work, we apply motion transfer on the Taobao product images in real e-commerce scenario to generate creative animations, which are more attractive than static images and they will bring more benefits. We animate the Taobao products of dolls, copper running horses and toy dinosaurs based on motion transfer method for demonstration.

## CCS CONCEPTS

• **Computing methodologies** → **Computer vision problems**; • **Applied computing** → **Online shopping**; • **Information systems** → **Multimedia content creation**.

## KEYWORDS

deep learning, image animation, motion transfer, e-commerce application

## 1 INTRODUCTION

Given a source image and a driving video of the same object, motion transfer based image animation aims at generating a result video preserving both the motion of the driving video and the appearance of the source image. Motion transfer has received increasing attention because of its great potential in practical applications, such as face swapping, dance transfer, virtual try-on and so on.

Previous methods on motion transfer only work on specific objects, such as human body[1] or human face[3]. Recently, a motion transfer method FOMM[2] is proposed, which works for arbitrary objects. However, existing works rarely transfer motion to objects other than human body or human face on public datasets.

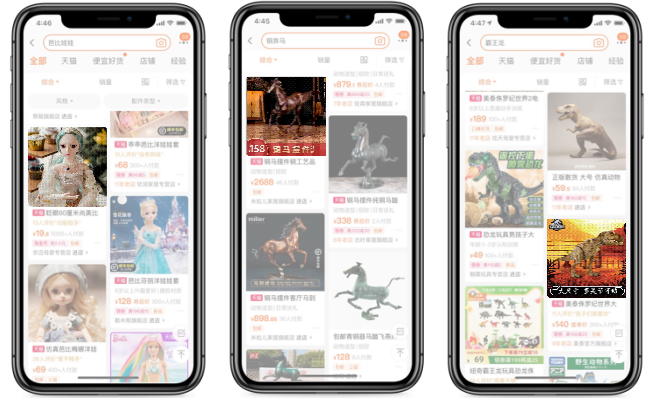

**Figure 1: Application demonstration in the Taobao e-commerce scenario. The highlighted parts are the animations generated by our system. We further demonstrate the animations in our supplementary video.**

In this paper, We apply motion transfer on the Taobao product images in real e-commerce scenario. For a Taobao product image, greater the attraction means greater the click, which also means greater the benefits. As shown in Figure 1, compared with a static image, an animation generated by motion transfer method is usually more attractive, especially when the moving object is supposed to be fixed. Therefore, creative animations generated via motion transfer are very helpful in e-commerce scenario.

Object categories include but are not limited to dolls, copper running horses and toy dinosaurs, whose Taobao product images are used for generating animation based on motion transfer method. To the best of our knowledge, we are the first work applying motion transfer on real e-commerce scenario. There are many potential difficulties from academic models to real applications. Our proposed system, which will be introduced in detail in the following chapters, solves these difficulties.

## 2 THE SYSTEM

The framework of our proposed system is shown in Figure 2. The system consists of three parts, pre-processing, motion transfer and post-processing.

### 2.1 Pre-Processing

The pre-processing part consists of two modules, object localization and background inpainting. The raw input of the system is a Taobao product image, which may contain multiple objects or small single object (human face in a full-body human image). However, the motion transfer module requires source image with single and big object on it.

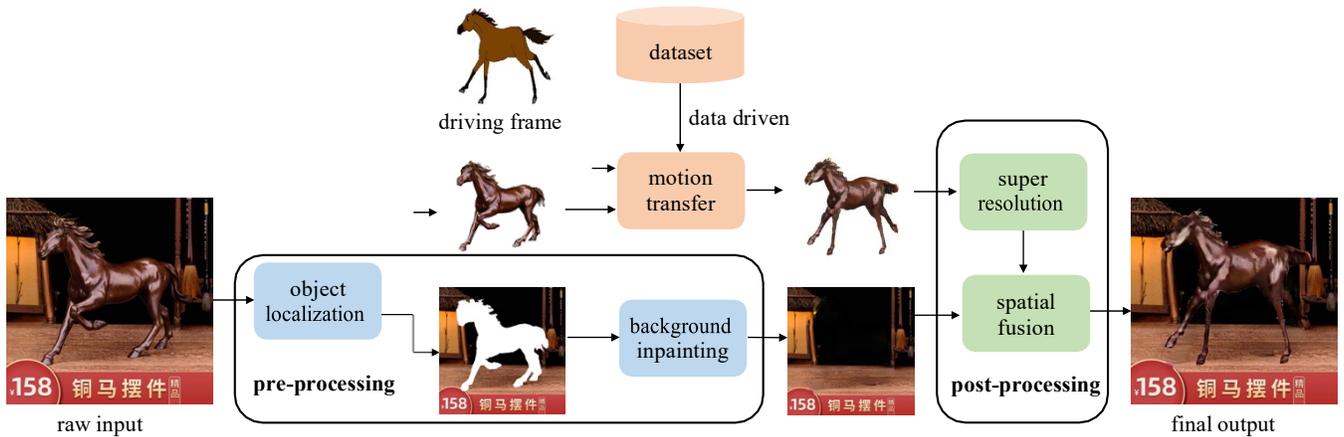

**Figure 2: Framework of our proposed system.**

Thus we introduce an *object localization* module to localize the object that we want to perform motion transfer on. This module can be an object detection model or a matting model[1], which separates foreground and background of the raw input for better animation quality. Object in the foreground is placed on a white background. However, the background is incomplete, so a *background inpainting* module is introduced to obtain a complete background. We use the CRA[4] inpainting model because of its good performance on large images.

## 2.2 Motion Transfer

We improve the motion transfer model FOMM[2] for better generated results, and we use the improved FOMM model as the *motion transfer* module. It consists of two models, motion model and generating model. The former extract and model the pose of the object, the latter generate image according to the extracted driving pose. Both models are based on encoder-decoder architecture. With a data driven manner, the *motion transfer* module is trained on a video dataset, which contains motion videos of the same kind of object as the source image. Different objects in source images need different datasets for model training.

The *motion transfer* module takes the pre-processed object image as source image and a frame from training videos as driving frame. Then the module will transfer the driving pose to the source image and generate a image with the appearance of the source image and the pose of the driving frame. Note that the animation is generated in a frame-by-frame manner. Moreover, object in the source image can make different motions via changing the driving video, which allows the object move as we like.

## 2.3 Post-Processing

The frame generated by the motion transfer module still needs post-processing with two modules, super resolution and spatial fusion. Firstly, a comparably large spatial size is needed for a Taobao product image but the spatial size of the generated frame is usually smaller. Enlarging it directly with upsampling will cause it to be blurred. Thus we introduce a *super resolution* module to enlarg the spatial size of the generated frame while ensuring its clarity.

---

[1] we use the matting model provided by Alimama and refine the result

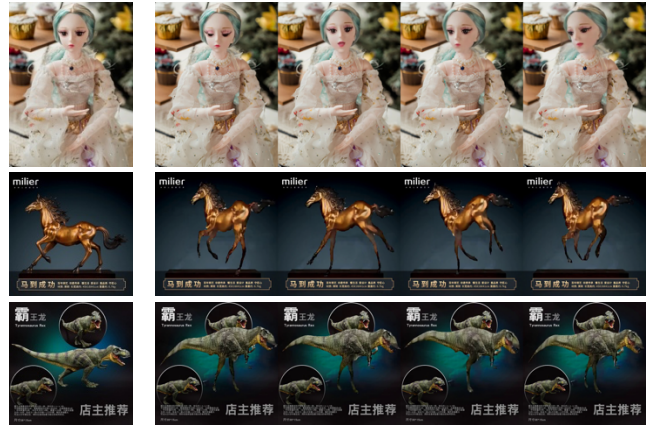

**Figure 3: Results of dolls, copper running horses and dinosaurs generated by our system. The left column is static product images, and the right four columns are the frames of the generated animations.**

The *spatial fusion* module overlays super-resolved generated frames on the pure background generated by the background inpainting module. Moreover, when stitching the animation back to the raw input image, we will fuse the edges of cropping to make the stitching softer.

Our final generated images are shown in Figure 3.

## 3 CONCLUSION

In this paper, we have proposed a system which can apply motion transfer on Taobao product images of arbitrary objects in e-commerce scenario. We generate creative and attractive animations of dolls, copper running horses and dinosaurs, which benefit a lot on Taobao advertisement scenario, based on the system for demonstration.